\documentclass[aps,twocolumn,prb,longbibliography,showpacs,floatfix,superscriptaddress]{revtex4-1}

\usepackage{graphicx,color}
\usepackage{amsfonts}
\usepackage[figuresright]{rotating}  
\usepackage{amssymb}
\usepackage{bm}
\usepackage{amsmath}
\usepackage{mathtools}
\usepackage{psfrag}
\usepackage{floatrow}
\usepackage{multirow}
\usepackage{tabularx}
\usepackage{textcomp}
\usepackage{units}
\usepackage{lipsum} 
\usepackage{soul}
\usepackage{titlesec}
\usepackage{times}
\usepackage{hyperref}
\usepackage{enumitem}
\usepackage{caption}
\usepackage{subcaption}
\DeclareMathAlphabet\mathbfcal{OMS}{cmsy}{b}{n}

\titlespacing*{\section}
{0pt}{1ex plus .25ex}{1ex plus 1ex}
\titlespacing*{\subsection}
{0pt}{1ex plus .25ex}{1ex plus 1ex}
\titlespacing*{\subsubsection}
{0pt}{1ex plus .25ex}{1ex plus 1ex}

\titleformat{\section}
 {\fontsize{10}{15}\bfseries }
  {\thesection}
  {1em}
  {}

\titleformat{\subsection}
  {\bfseries }
  {\thesubsection}
  {2em}
  {}

\titleformat{\subsubsection}
  {\itshape}
  {\thesubsubsection}
  {2em}
  {}

\def\beq{\begin{eqnarray}}
\def\eeq{\end{eqnarray}}

\newcommand{\ket}[1]{\left| #1 \right>} 
\newcommand{\bra}[1]{\left< #1 \right|} 
\newcommand{\braket}[2]{\left< #1 \vphantom{#2} \right|
 \left. #2 \vphantom{#1} \right>} 
\let\baraccent=\= 
\renewcommand{\=}[1]{\stackrel{#1}{=}} 
\newcommand{\mc}[1]{\mathcal{ #1}} 

\makeatletter

\titleclass{\subsubsubsection}{straight}[\subsection]

\newcounter{subsubsubsection}[subsubsection]
\renewcommand\thesubsubsubsection{\thesubsubsection.\arabic{subsubsubsection}}

\titleformat{\subsubsubsection}
   {\normalfont\normalsize\itshape \centering}{\thesubsubsubsection}{1em}{}
\titlespacing*{\subsubsubsection}
{0pt}{1ex plus .3ex}{1ex plus .3ex}

\makeatletter
\renewcommand\paragraph{\@startsection{paragraph}{5}{\z@}%
  {3.25ex \@plus1ex \@minus.2ex}%
  {-1em}%
  {\normalfont\normalsize}}
\renewcommand\subparagraph{\@startsection{subparagraph}{6}{\parindent}%
  {3.25ex \@plus1ex \@minus .2ex}%
  {-1em}%
  {\normalfont\normalsize}}
\def\toclevel@subsubsubsection{4}
\def\toclevel@paragraph{5}
\def\toclevel@paragraph{6}
\def\l@subsubsubsection{\@dottedtocline{4}{7em}{4em}}
\def\l@paragraph{\@dottedtocline{5}{10em}{5em}}
\def\l@subparagraph{\@dottedtocline{6}{14em}{6em}}
\makeatother

\newcommand{\ii}{\mathrm{i}}
\newcommand{\ee}{\mathrm{e}}

\begin{document}
\title{Crystalline finite-size topology}
\author{Micha\l{} J. Pacholski}
\affiliation{Max Planck Institute for Chemical Physics of Solids, Nöthnitzer Strasse 40, 01187 Dresden, Germany}
\affiliation{Max Planck Institute for the Physics of Complex Systems, Nöthnitzer Strasse 38, 01187 Dresden, Germany}

\author{Ashley M. Cook}
\affiliation{Max Planck Institute for Chemical Physics of Solids, Nöthnitzer Strasse 40, 01187 Dresden, Germany}
\affiliation{Max Planck Institute for the Physics of Complex Systems, Nöthnitzer Strasse 38, 01187 Dresden, Germany}

\begin{abstract}
Topological phases stabilized by crystalline point group symmetry protection are a large class of symmetry-protected topological phases subjected to considerable experimental scrutiny. Here, we show that the canonical three-dimensional (3D) crystalline topological insulator protected by time-reversal symmetry $\mc{T}$ and four-fold rotation symmetry $\mc{C}_4$ individually or the product symmetry $\mc{C}_4 \mc{T}$,  generically realizes finite-size crystalline topological phases in thin film geometry (a quasi-(3-1)-dimensional, or q(3-1)D, geometry): response signatures of the 3D bulk topology co-exist with topologically-protected, quasi-(3-2)D and quasi-(3-3)D boundary modes within the energy gap resulting from strong hybridisation of the Dirac cone surface states of the underlying 3D crystalline topological phase. Importantly, we find qualitative distinctions between these gapless boundary modes and those of strictly 2D crystalline topological states with the same symmetry-protection, and develop a low-energy, analytical theory of the finite-size topological magnetoelectric response.
\end{abstract}
\maketitle

Crystalline topological phases, or those protected in whole or in part by crystalline point group symmetries, have been a very active front in efforts to identify and classify topologically non-trivial phases of matter. The large number of crystalline point group symmetries protect many distinctive topological insulator and semimetal states~\cite{fu2011cryst, Benalcazar61, tanaka2012, weng2014, ando2015, Wieder2018, PhysRevB.85.165120, PhysRevB.83.245132, PhysRevB.86.115112, PhysRevB.82.241102, PhysRevB.91.155120, PhysRevB.93.195413, PhysRevB.95.235425, Aroyo:xo5013, slager_space_2013, PhysRevX.7.041069, doi:10.1126/sciadv.1501782, PhysRevB.93.205104, doi:10.1073/pnas.1514665112, PhysRevLett.117.096404, bradlyn_topological_2017, PhysRevB.96.035115}, building extensively on the foundational work of the ten-fold way classification scheme~\cite{Ryu_2010, Schnyder2008}. Recent work reveals, however, that these canonical D-dimensional states, such as the Chern insulator~\cite{haldane1988}, or the strong topological insulator~\cite{Moore_2007}, can remain relevant even when the system is only thermodynamically large in $\delta < D$ directions~\cite{cookFST_2023, calderon2023}: for example, taking $\delta=1$, even if $(D-1)$-dimensional gapless boundary modes associated with a D-dimensional bulk topological invariant are lost due to strong hybridisation, D-dimensional topological response signatures can co-exist with quasi-(D-2)-dimensional (q(D-2)D) gapless boundary modes in the form of \textit{finite-size topological phases}~\cite{cookFST_2023}. $N$-fold degrees of freedom, with $1<N<10$, can then potentially serve as synthetic dimensions, greatly enriching physics of band topology.

\begin{figure}[htb!]
\includegraphics[width=1\columnwidth]{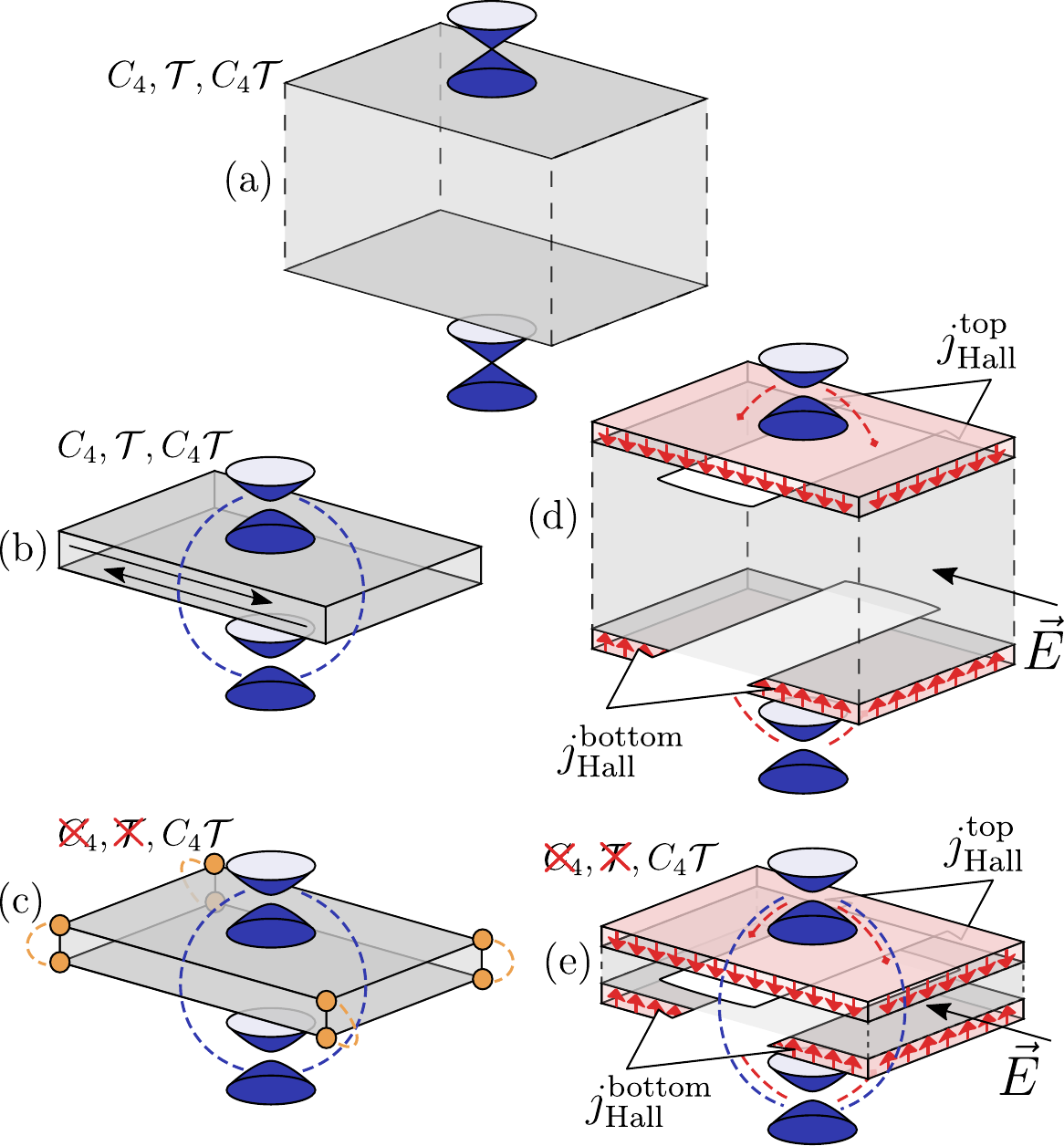}
\caption{Demonstrating crystalline finite-size topology: a) 3D crystalline topological insulator with Dirac cone surface states. b) and c) depict bulk-boundary correspondence of system in thin-film geometry either with time-reversal symmetry (TRS) and four-fold rotation symmetry ($\mc{C}_4$) individually present or without, respectively, corresponding to finite-size topological phase. d) 3D system with magnetic surface perturbations to probe quantised surface Hall conductivity associated with the 3D topological state, e) system in thin film geometry realising bulk-boundary correspondences of b) and c), with magnetic surface perturbations to confirm topological magnetoelectric response of finite-size topological phase.\label{fig1}}
\end{figure}

In this work, we demonstrate finite-size topological phases are realized for crystalline topological insulators as well. We focus on the canonical Hamiltonian for the first formally-identified crystalline topological phase~\cite{fu2011cryst}, a $3$D topological insulator protected by four-fold rotational symmetry and time-reversal symmetry. That is, we confirm that a system realizing the canonical crystalline topological state in the 3D bulk, but which is thermodynamically large in only two spatial dimensions, realizes quasi-($3-3$)D gapless boundary corner states or quasi-($3-2$)D gapless boundary edge states when $2$D gapless boundary modes of the 3D phase strongly hybridise, \textit{while still possessing the topological response signature of the 3D bulk invariant.} The system geometries and procedures for confirming these two defining properties of finite-size topological phases are shown schematically in Fig.~\ref{fig1}. Our work therefore lays the foundation for far broader study of topological phases protected in whole or in part by crystalline point group symmetries, with the foundational results presented here particularly important in understanding of Van der Waals thin films and heterostructures\cite{vanwaals-hetero,Nature-hetero-antiferr,TI-hetero,robust-2dhetero, TMD-1,TMD-topo, Song_2018, qian_2014, kawamura_laughlin_2023} identified as hosting 2D or quasi-1D topological states, which may actually be partially-identified finite-size topological phases instead descending from underlying higher-dimensional bulk topology.

\textit{Hamiltonian}---We consider a Hamiltonian  previously-introduced by Fu~\cite{fu2011cryst} realizing the crystalline topological insulator phase. The Bloch Hamiltonian is taken to be
\begin{align}
&H(\bm k) = \Big(M+t\sum_{i \in\{x,y,z\}}\cos k_i\Big)\tau_z\sigma_0 + \Delta_1\tau_x \bm\sigma\cdot \sin \bm k \nonumber \\ &\qquad + (\cos k_x - \cos k_y)(\Delta_2\tau_y \sigma_0 + \Delta_3 \tau_0\sigma_z)
\,.
\end{align}
At $\Delta_2=\Delta_3=0$, the Hamiltonian respects both time-reversal symmetry $\mathcal{T} = \sigma_y \mathcal{K}$ and four-fold rotation symmetry
\begin{equation}
  C_4 = \ee^{\ii(\sigma_z/2 + xk_y - y k_x)\pi/2}\,,
\end{equation}
which acts as
\begin{equation}
\begin{gathered}
  C_4^\dag (k_x,k_y,k_z) C_4 = (k_y,-k_x,k_z)\,, \\
  C_4^\dag (\sigma_x,\sigma_y,\sigma_z) C_4 = (\sigma_y,-\sigma_x,\sigma_z)\,.
\end{gathered}
\end{equation}
The terms proportional to $\Delta_2,\Delta_3$ are the simplest such terms (i.e. containing only nearest-neighbor hoppings) that break both $\mathcal{T}$ and $C_4$ symmetries, while preserving their product $C_4\mathcal{T}$. 

\textit{Phase diagram}---We are interested in characterizing the phase diagram of this model, in particular in a finite-thickness slab geometry, and its properties that generalize to arbitrary systems in the class of $C_4\mathcal{T}$-symmetric crystalline topological insulators. To do so, we first briefly review standard characterization for the system thermodynamically large in three space dimensions. In three-dimensions, the model is characterized by a $\mathbb{Z}_2$ invariant, which distinguishes phases with and without gapless Dirac cones at the $C_4\mathcal{T}$-invariant surfaces (i.e. $z=\text{const}$). 

We can demonstrate this bulk-boundary correspondence by calculating the surface states of our toy model. We know that the Dirac point will be located at a $C_4\mathcal{T}$-invariant surface momentum $(k_x,k_y) = q\in\{(0,0),(\pi,\pi)\}$, and, due to the presence of an additional artificial particle-hole symmetry $\mathcal{C} = \tau_y \sigma_y \mathcal{K}$, the surface-gap closing will occur at energy $E=0$, and the zero-mode will be an eigenstate of chiral symmetry operator $\mathcal{T}\mathcal{C} = \tau_y$: $\tau_y\psi = \chi\psi$, $\chi=\pm 1$. The gap closing condition then reads
\begin{equation}
\left(m_q+t\cos k_z + \ii \chi \Delta_1 \sigma_z \sin k_z\right) \psi= 0\,,
\end{equation}
where the $m_{0,0} = M+2t$, $m_{\pi,\pi} = M-2t$. Solving for $k_z$, we find
\begin{equation}
\ee^{\ii k_z} = \frac{-m_q \pm \sqrt{m_q^2 + \Delta_1^2 - t^2}}{t + s\chi \Delta_1}\,.
\end{equation}
with $s=\pm 1$, $\sigma_z\psi = s\psi$.

For a state with given $\chi,s$ to be decaying at ${z\rightarrow\infty}$, there must be two solutions for $k_z$, satisfying ${|\ee^{\ii k_z}| < 1}$. This is true if and only if $\chi s = \operatorname{sign}(\Delta_1/t)$ and $|m_q|<|t|$.

If cones are present at both $q=(0,0)$ and $q=(\pi,\pi)$ simultaneously, they are no longer protected, which results in a trivial phase. This leads to a nontrivial bulk-topological phase for $-3|t|<M<-|t|$ and $2|t|<M<3|t|$.

\textit{Quasi-(3-1)D thin-film geometry}---Now, we will turn our attention to a slab of thickness $L$ finite in $z$ direction. In a topological region of the 3D bulk phase diagram, the overlap between the surface states on the two surfaces will produce a surface hybridisation gap, which may oscillate with the slab thickness and the parameters of the model.

\begin{figure}[htb!]
    \centering
    \includegraphics[width=\linewidth]{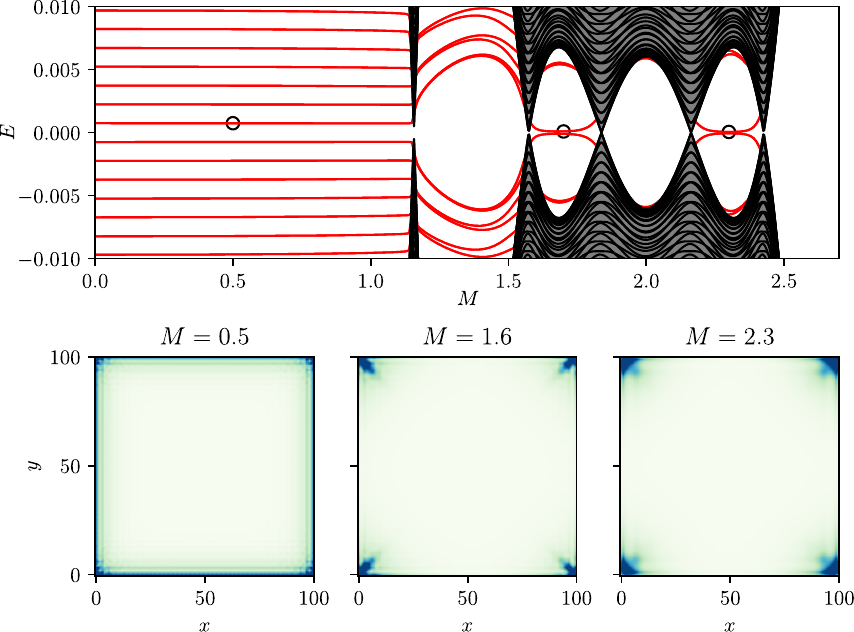}
    \caption{Bulk-boundary correspondence of finite-size crystalline topological phase: a) energy spectrum vs. mass term $M$ of system in thin film geometry with OBC in the $\hat{z}$-direction (black) and OBC in each spatial direction (red). b), c) and d) depict charge density distribution vs. $x$ and $y$ for states highlighted by black circles in a) corresponding to $M = 0.5$, $1.6$, and $2.3$, respectively. The values of the remaining parameters used were  $t=-1$, $\Delta_1=0.6$,  $\Delta_2=0.37$, $\Delta_3=0.4$.}
    \label{fig2}
\end{figure}

We can capture this phenomenology with a low-energy model of the surface states. Assuming that the slab spans $0<z\leq L$, we know that the surface states at $z=0$ will have $\chi s=\operatorname{sign}(\Delta_1/t)$, whereas those at $z=L$ will have $\chi s=-\operatorname{sign}(\Delta_1/t)$. For a surface momentum close to the Dirac point $(k_x,k_y) = q + \delta q$, we can project the Hamiltonian onto the subspace spanned by the surface states.
Taking for concreteness $t=1$, $\Delta_1>0$, we then get the low energy effective Hamiltonian (additional details of derivation provided in the SM~\cite{sup})
\begin{align}
&h = \Delta_1 \nu_0 (\sigma_x \delta q_x + \sigma_y \delta q_y) + \nu_y \sigma_z \delta \nonumber \\
&\qquad + s_q (\delta q_x^2 - \delta q_y^2)(\Delta_2 \nu_z\sigma_z + \Delta_3 \nu_0\sigma_z)
\label{eq:surfham}
\,,
\end{align}
where $\nu_\alpha$ is a set of Pauli matrices acting in the surface-index degree of freedom, $\delta$ is the hybridisation gap, and $s_{0,0}=-1$, $s_{\pi,\pi}=1$.

Let us now consider a slab finite in $x,y$ directions (with size $W\gg L$). The vacuum can be modeled by taking $\delta\rightarrow \infty$. If time-reversal symmetry is preserved ($\Delta_2=\Delta_3=0$), and $\delta<0$ in the interior of the slab, we expect the edge states propagating along the edges. We shall find them analytically. For an edge along $\bm n_\parallel = (\cos\alpha, \sin\alpha)$, with vacuum at $\bm r\cdot \bm n_\perp < 0$, $\bm n_\perp=\bm n_\parallel \times \hat{\bm z}=(\sin\alpha, -\cos\alpha)$ is a unit vector pointing towards the bulk of the slab, the boundary condition is $\nu_y (\bm n_\parallel\cdot \bm \sigma)\psi = -\psi$. The edge-state solutions are then
\begin{equation}
    \psi_\pm \propto \ee^{-(\delta/\Delta_1) \bm r \cdot \bm n_\perp} \ee^{\ii \delta q_\parallel \bm r \cdot \bm n_\parallel} \binom{1}{\pm \ii}_\nu \binom{1}{\mp\ee^{\ii\alpha}}_\sigma\ ,
\end{equation}
with  corresponding energies
\begin{equation}
    E_\pm(q_\parallel) = \mp \Delta_1 \delta q_\parallel\,,
\end{equation}
where $\delta q_\parallel$ is the momentum along the edge. Projecting the low-energy slab Hamiltonian onto the edge-state Hilbert space---this time allowing for non-zero $\Delta_2,\Delta_3$---we get a low-energy edge Hamiltonian
\begin{equation}
    h' = -\Delta_1 q_\parallel \nu_y + s_q \frac{\Delta_2 \delta^2}{\Delta_1^2} \cos (2\alpha)  \nu_z\,.
\end{equation}
The mass term proportional to $\nu_z$ changes sign at the points where edge orientation is $\alpha\in\{\pi/4, 3\pi/4, 5\pi/4, 7\pi/4\}$, at which points there will be corner zero-energy bound states. These corner states will be eigenstates of $\nu_x$, so their charge distribution will be equally split between the top and the bottom surface.

\begin{figure}[htb!]
    \includegraphics[width=       \linewidth]{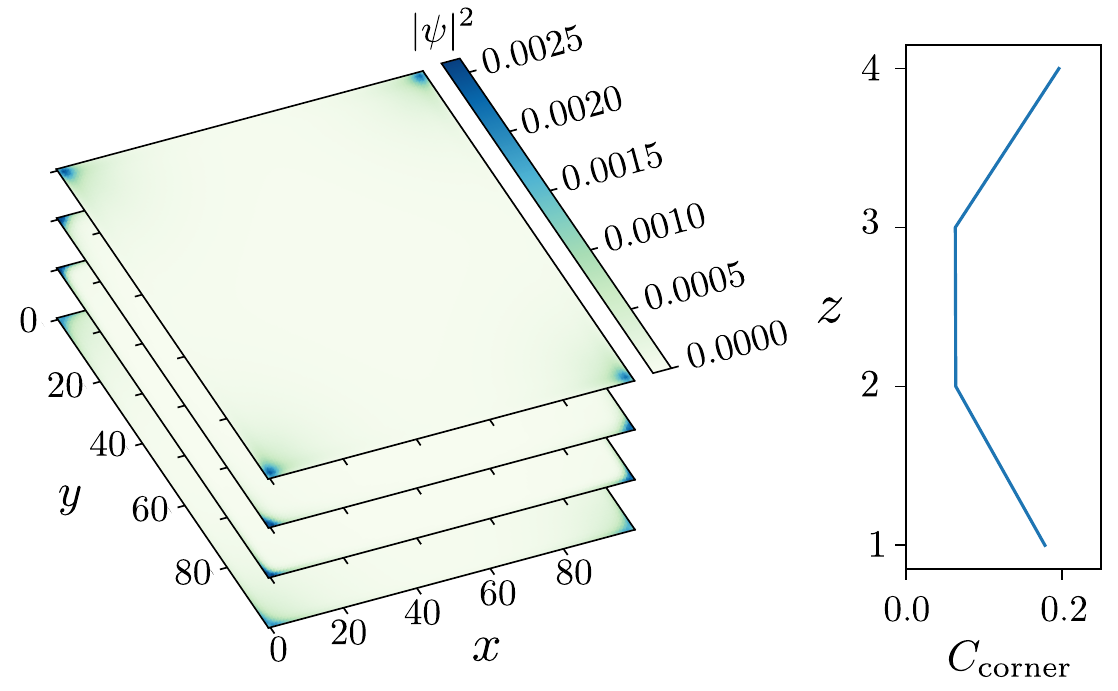}
    \caption{Left panel: probability density distribution over real-space for corner mode shown in Fig.~\ref{fig2} at $M=1.6$, with four unit cells in the stacking ($z$-) direction. Right panel: charge density per corner vs. layer index in the stacking ($z$-) direction.}
    \label{fig:cornermode}
\end{figure}

We support the above analytical calculations with numerical results for the quasi-(3-1)D slab shown in Fig.~\ref{fig2}, with open-boundary conditions in first the $z$ direction (black) and then also in the $x$ and $y$ directions (red). We characterize the quasi-(3-1)D bulk topology with the topological invariant $\nu$~\cite{day2023}
\begin{equation}
    \nu = {1 \over \pi } \left[ \int_{\mathrm{IBZ}} \mathrm{Tr} \mathcal{F} d\boldsymbol{k}^2 + 2 i \log \widetilde{\mathrm{det}}\mathcal{W}_{\Gamma \rightarrow M}\right] \mod 4,
\end{equation}
defined over the irreducible Brillouin zone (IBZ), where $\mathcal{F}$ is the non-Abelian Berry curvature and $\widetilde{\mathrm{det}}\mathcal{W}_{\mathcal{C}}$ is the dressed Wilson line determinant.

In the case when $\Delta_2=0$, this invariant can be calculated explicitly for the effective low energy Hamiltonian~\cite{sup}:
\begin{align}
\nu
&= 
1 - \operatorname{sign}(\delta_{0,0}\delta_{\pi,\pi}) 
- \operatorname{sign}(2 \Delta_3 + \delta_{0,\pi}) \notag
\\
&\qquad  + \operatorname{sign}(2 \Delta_3 - \delta_{0,\pi})
\mod 4\,.
\end{align}

When $\mathcal{T}$ and $\mathcal{C}_4$ are present $(\Delta_2 = \Delta_3 = 0)$, $\nu$ is $\mathbb{Z}_2$-classified, and $\mathbb{Z}_4$ when these symmetries are broken while preserving $\mathcal{C}_4 \mathcal{T}$. As shown in Fig.~\ref{fig2}, non-trivial $\nu$ corresponds to quasi-(3-2)D gapless edge states ($\nu = 1,3$) or quasi-(3-3)D corner modes ($\nu = 2$) for this geometry.

It is important to explicitly distinguish between corner states of a strictly 2D topological phase and the corner states of the finite-size topological phase presented here. The probability density distribution of boundary states in the finite-size topological phase are noticeably $z$-dependent as shown in Fig.~\ref{fig:cornermode}, with charge density concentrated at the corners of the top and bottom layers specifically, rather than evenly distributed along the hinges. This bulk-boundary correspondence distinguishes the finite-size topological phase from a strictly 2D crystalline topological state.

\textit{Topological response signatures of finite-size topology}---We may also examine the topological response of the system normally associated with a 3D bulk, the topological magnetoelectric polarizability~\cite{essin2009}, for the system in the quasi-(3-1)D geometry, to further investigate the nature of the topological non-trivial state. To do so, we introduce magnetic perturbations at the top and bottom surfaces of the quasi-(3-1)D system as illustrated in Fig.~\ref{fig1}(e).

\begin{figure}[htb!]
    \centering
    \includegraphics[width=\linewidth]{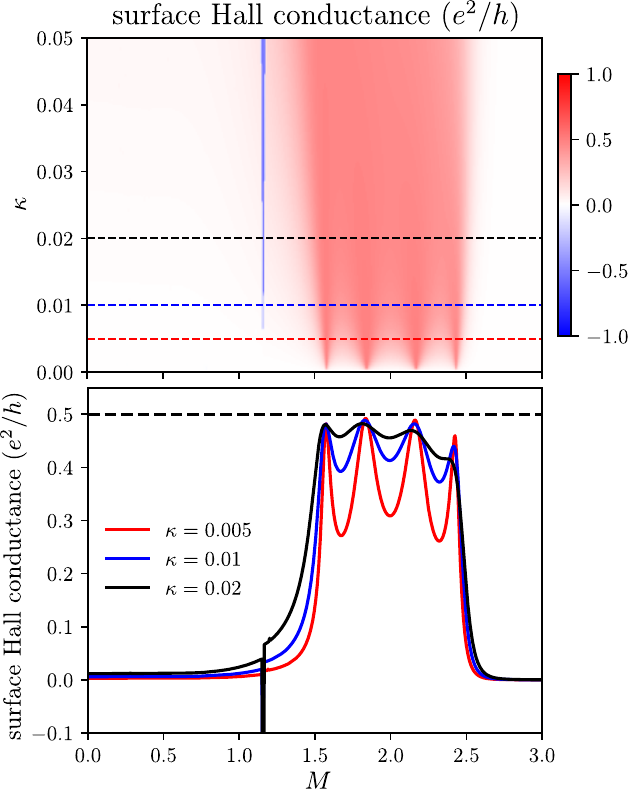}
    \caption{Topological magnetoelectric response of finite-size crystalline topological phase: a) surface Hall conductance vs. mass term $M$ and surface magnetization strength $\kappa$. b) depicts cuts through a) for different fixed magnetisation strengths $\kappa$. }
    \label{fig3}
\end{figure}

We model the magnetic perturbation by adding a term $\tilde\kappa \nu_z\sigma_z$ to the effective surface Hamiltonian \eqref{eq:surfham}.
\begin{equation}
h = \Delta_1 \nu_0 (\sigma_x \delta q_x + \sigma_y \delta q_y) + \nu_y\sigma_z \delta + \tilde\kappa \nu_z\sigma_z
\end{equation}
The Hall conductivity of the top/bottom surfaces is given by the formula~\cite{essin2009}
\begin{equation}
C_\pm = \frac{\ii}{2\pi}\int d k \mathrm{Tr} [P \epsilon_{ij} (\partial_i P) \nu_\pm (\partial_j P)]\,
\end{equation}
where $P$ is the ground-state projector, and $\nu_\pm = \frac{1}{2}(\nu_0 \pm \nu_z)$ is the projector onto top/bottom surface. We can find the spectrum of the Hamiltonian by squaring it
\begin{equation}
H^2 = \epsilon^2\,, \qquad \epsilon = \sqrt{v^2 k_x^2 + v^2 k_y^2 + \delta^2 + \tilde\kappa^2}\,,
\end{equation}
thus the ground-state projector is simply
\begin{equation}
P = \frac{\epsilon-H}{2\epsilon}
\end{equation}
We then get
\begin{equation}
\mathrm{Tr} [P \epsilon_{ij} (\partial_i P) \nu_\pm (\partial_j P)] =
\mp \frac{\ii\tilde\kappa}{2\epsilon^3}
\end{equation}
which yields
\begin{equation}
C_\pm = \pm\frac{\tilde\kappa}{2\sqrt{\delta^2 +\tilde\kappa^2}}\,,
\end{equation}
In the case when the magnetization-induced gap dominates over the hybridisation gap, this tends to the expected value $1/2$. These results may provide additional understanding of past work on the magnetoelectric polarisability of axion insulators in thin film systems, where deviations from $1/2$ are also observed as part of the topological response signature~\cite{kawamura_laughlin_2023}

Numerical results on the response signatures associated with the topological magnetoelectric polarizability are shown in Fig.~\ref{fig3}. Surface Hall conductance for the top layer is shown over the same interval in mass parameter $M$ as in Fig.~\ref{fig2} as a function of magnetization strength $\kappa$. We see that, with increasing $\kappa$, the surface Hall conductance increasingly approaches a saturation value of $0.5$ in units of $e^2 / h$, the value associated with non-trivial magnetoelectric polarizability~\cite{essin2009}, over the region of underlying bulk 3D topological state. For very small $\kappa$, finite surface Hall conductance first nucleates about the transition points between topological bubbles corresponding to different finite-size topological states, competing with the hybridisation gap. This demonstrates that, even if the gapless surface states associated with the 3D bulk topological phase strongly hybridise and are lost, in this sense, due to finite-size effects, the 3D bulk topological invariant remains very relevant in characterizing the topological state, and the quasi-(3-1)D system is not adequately characterized by topological invariants of strictly 2D, 1D and/or 0D bulk. This may help explain recent experiments in thin-film systems investigating states with non-trivial magnetoelectric polarisability~\cite{liu_robust_2020, kawamura_laughlin_2023, li2023, liu_magnetic-field-induced_2021}.

\textit{Discussion \& Conclusion}---We introduce crystalline finite-size topological phases of matter in this work. We examine the Hamiltonian of the canonical crystalline topological insulator state protected by four-fold rotational symmetry and time-reversal symmetry, or invariance of the system under the product operation of four-fold rotation and time-reversal~\cite{fu2011cryst}. For open boundary conditions in the $z$ direction, with corresponding system size in this direction, $L$, on the scale of a few unit cells (e.g., $L<10$), we find the gapless surface states occurring for thermodynamically large $L$ generically strongly hybridise to open a gap, with gapless regions reduced to gapless, fine-tuned \textit{transition} points between topologically-distinct gapped regions of the phase diagram. These gapped regions may be topologically-characterized to determine an additional bulk-boundary correspondence distinct from that of strictly 2D topological states, with non-trivial invariant indicating gapless edge or corner states concentrated at the top and bottom surfaces upon opening boundary conditions in the $x$ and $y$ directions such that the protecting symmetries of the bulk state are preserved at the boundary. As required for a finite-size topological phase, however, we also confirm the layer-dependent Hall conductance signature of non-trivial magnetoelectric polarizability persists for $L<10$ even when the surface states of the underlying 3D state are absent due to strong hybridisation.\par
Our work therefore serves as a foundation in studying finite-size topology of the large class of topological states protected in whole or in part by crystalline point group symmetries and studied heavily in experiments. Our work may furthermore provide understanding of previously-observed topological response signatures of intrinsically three-dimensional topological states observed in thin film systems~\cite{liu_robust_2020, kawamura_laughlin_2023, li2023, liu_magnetic-field-induced_2021}.

\textit{Acknowledgements}---This research was supported in part by the National Science Foundation under Grants No.NSF PHY-1748958 and PHY-2309135, and undertaken in part at Aspen Center for Physics, which is supported by National Science Foundation grant PHY-2210452.

\bibliography{ref.bib}

\cleardoublepage

\vspace{4in}


\clearpage

\makeatletter
\renewcommand{\theequation}{S\arabic{equation}}
\renewcommand{\thefigure}{S\arabic{figure}}
\renewcommand{\thesection}{S\arabic{section}}
\setcounter{equation}{0}
\setcounter{section}{0}
\onecolumngrid
\begin{center}
  \textbf{\large Supplemental material for ``Crystalline finite-size topology''}\\[.2cm]
  Micha\l{} J. Pacholski,$^{1,2}$ and Ashley M. Cook$^{1,2,*}$\\[.1cm]
  {\itshape ${}^1$Max Planck Institute for Chemical Physics of Solids, Nöthnitzer Strasse 40, 01187 Dresden, Germany\\
  ${}^2$Max Planck Institute for the Physics of Complex Systems, Nöthnitzer Strasse 38, 01187 Dresden, Germany\\}
  ${}^*$Electronic address: cooka@pks.mpg.de\\
(Dated: \today)\\[1cm]
\end{center}

\section{Derivation of the low-energy effective Hamiltonian}

The Hamiltonian we use in the main text.
\begin{equation}\label{eq:ham}
H(\bm k) = \Big(M+t\sum_{i=1}^3\cos k_i\Big)\tau_z\sigma_0 + \Delta_1\tau_x \bm\sigma\cdot \sin \bm k \nonumber + (\cos k_x - \cos k_y)(\Delta_2\tau_y \sigma_0 + \Delta_3 \tau_0\sigma_z)
\,.
\end{equation}
which has $C_4\mathcal{T} = \ee^{-\ii\sigma_z\pi/4}\sigma_y\mathcal{K}$ symmetry:
\begin{equation}
C_4\mathcal{T} H(k_x,k_y) (C_4\mathcal{T})^{-1} = H(-k_y,k_x)\,.
\end{equation}
For now we'll set $\Delta_2=\Delta_3=0$, which restores time-reversal symmetry $\mathcal{T} = \sigma_y\mathcal{K}$. Surface Dirac cones can occur at time-reversal invariant momenta (TRIMs) $(k_x,k_y)=q \in ((0,0),(\pi,\pi), (\pi,0),(0,\pi))$. At these momenta, the Hamiltonian becomes
\begin{equation}
H(q) = (m_q + t\cos k_z)\tau_z\sigma_0 + \Delta_1\tau_x \sigma_z \sin k_z \,.
\end{equation}
where
\begin{equation}
m_{0,0} = M+2t\,,\qquad m_{\pi,\pi}=M-2t\,,\qquad m_{0,\pi}=m_{\pi,0}=M\,.
\end{equation}
The Hamiltonian commutes with $\sigma_z$ and anticommutes with $\tau_y$, thus the zero modes will take form
\begin{equation}
\ket\psi = \frac{1}{\sqrt{2}}\binom{1}{\ii\chi}_\tau\ket{\sigma_z=s} \sum_z \zeta^z \ket{z}\,,\qquad
\chi,s=\pm 1
\end{equation}
where $\zeta \equiv \ee^{\ii k_z}$. The eigenequation takes form
\begin{equation}
\left(m_i+t\frac{\zeta+\zeta^{-1}}{2}\right) + \ii \chi s\Delta_1 \frac{\zeta - \zeta^{-1}}{2\ii} = 0\,.
\end{equation}
The solutions for $\zeta$ are
\begin{equation}
\zeta_\pm = \frac{-m_i \pm \sqrt{m_i^2 - t^2 + \Delta_1^2}}{t + \chi s \Delta_1}
\end{equation}
An eigenstate of a semi-infinite system in $z$ direction, must satisfy the boundary condition $\psi(z=0) = 0$. This is possible if and only if $|\zeta_+|,|\zeta_-|<1$ for the $z>0$ surface, and $|\zeta_+|,|\zeta_-|>1$ for the $z<0$ interface. Identity
\begin{equation}
\frac{1}{\zeta_\pm(\chi s=1)} = \frac{t + \Delta_1}{-m_i \pm \sqrt{m_i^2 - t^2 + \Delta_1^2}} 
=
\frac{-m_i \mp \sqrt{m_i^2 - t^2 + \Delta_1^2}}{t-\Delta_1} = \zeta_\mp(\chi s=-1)\,.
\end{equation}
ensures that if the condition is satisfied for one surface for given value of $\chi s$, then its automatically satisfied for the other surface, with opposite value of $\chi s$.

If the condition is satisfied, the eigenstate is a superposition
\begin{equation}
\ket{\psi} = \mathcal{N}
\frac{1}{\sqrt{2}}\binom{1}{\ii\chi}_\tau\ket{\sigma_z=s} \sum_z (\zeta_+^z-\zeta_-^z) \ket{z}.
\end{equation}
Let's focus on the condition $|\zeta_+|,|\zeta_-|<1$. It is satisfied for $\chi s = \mathrm{sign}(\Delta_1/t)$ for $|m_i|<|t|$. Suppose that $\zeta_+,\zeta_-$: $|\zeta_+|,|\zeta_-|< 1$ are solution for $\chi s=1$, i.e. $(\chi,s) \in \{ (1,1), (-1,-1) \}$, which results in a pair of surface states at one of two surfaces ($z=1$). Then around $k=q$ the low-energy surface subspace is spanned by states
\begin{equation}
\ket{\psi_+} = \underbrace{\frac{1}{\sqrt{2}} \binom{1}{\ii}}_{\ket{\tau_y=1}}\ket{\sigma_z = 1} \ket{\varphi}\,,\qquad
\ket{\psi_-} = \underbrace{\frac{1}{\sqrt{2}} \binom{\ii}{1}}_{\ket{\tau_y=-1}}\ket{\sigma_z = -1} \ket{\varphi}\,,\qquad
\end{equation}
where $\ket{\varphi} = \mathcal{N}\sum_{z=1}^L (\zeta_+^z - \zeta_-^z) \ket{z}$ is the normalized spacial part of the wavefunction. The Hamiltonian projected onto this subspace reads
\begin{equation}
H_\text{surf} = 
\begin{pmatrix}
\bra{\psi_+}H\ket{\psi_+} & \bra{\psi_+}H\ket{\psi_-} \\
\bra{\psi_-}H\ket{\psi_+} & \bra{\psi_-}H\ket{\psi_-}
\end{pmatrix}
=
\begin{pmatrix}
0 & \Delta_1(k_x - \ii k_y) \\
\Delta_1(k_x + \ii  k_y) & 0
\end{pmatrix}
= \Delta_1 (\sigma_x k_x + \sigma_y k_y)
\end{equation}
On the opposite surface (at $z=L$), states with $\chi s = -1$ are solutions:
\begin{equation}
\ket{\psi_+'} = \frac{1}{\sqrt{2}} \binom{\ii}{1}\ket{\sigma_z = 1} \ket{\varphi'}\,,\qquad
\ket{\psi_-'} = \frac{1}{\sqrt{2}} \binom{1}{\ii}\ket{\sigma_z = -1} \ket{\varphi'}\,,\qquad
\end{equation}
where $\ket{\varphi'} = \mathcal{N}\sum_{z=1}^L (\zeta_+^{L+1-z} - \zeta_-^{L+1-z}) \ket{z}$, in the basis of which the  low-energy Hamiltonian is the same as for the first surface:
\begin{equation}
H_\text{surf}'
= \Delta_1 (\sigma_x k_x + \sigma_y k_y)
\end{equation}
Introducing a Pauli matrix $\nu_z$, which labels the two surfaces, we can write the low-energy Hamiltonian as
\begin{equation}
h = \Delta_1 \nu_0 (\sigma_x k_x + \sigma_y k_y)
\end{equation}
In the first approximation, we can find the hybrydization gap, by calculating the overlap terms
\begin{align}
\bra{\psi_\alpha}H\ket{\psi_\beta'} &= 
\braket{\psi_\alpha}{z=1}\bra{z=1}H\ket{\psi_\beta'} \nonumber \\
&=
\bra{\tau_y=\alpha}\bra{\sigma_z=\alpha}
(\zeta_+ - \zeta_-)
\bra{z=1}H\ket{\varphi'}
\ket{\tau_y=-\beta}\ket{\sigma_z=\beta} \nonumber  \\
&=
\bra{\tau_y=\alpha}\bra{\sigma_z=\alpha}
\varphi^*(1)
[(m_i \varphi'(1) + t\frac{1}{2}\varphi'(2))\tau_z\sigma_0 + \Delta_1\tau_x \sigma_z \frac{1}{2\ii} \varphi'(2)]
\ket{\tau_y=-\beta}\ket{\sigma_z=\beta} \nonumber  \\
&=
\varphi^*(1)
[
\ii\alpha\delta_{\alpha\beta}
(m_i \varphi'(1) + t\frac{1}{2}\varphi'(2))
+ 
\alpha\delta_{\alpha\beta}
\Delta_1 \frac{1}{2\ii} \varphi'(2)] \nonumber 
\\
&=
\ii\alpha\delta_{\alpha\beta}
\varphi^*(1)
[
\underbrace{m_i \varphi'(1)
+ t\frac{1}{2}(\varphi'(2)+\varphi'(0))
-\Delta_1 \frac{1}{2} (\varphi'(2)-\varphi'(0))}_{=0}
- t\frac{1}{2}\varphi'(0)
-\Delta_1 \frac{1}{2} \varphi'(0) \nonumber 
] \\
&=
-\ii\alpha\delta_{\alpha\beta}
\frac{t+\Delta_1}{2} 
\varphi^*(1)\varphi'(0) \nonumber  \\
&=
-\ii\alpha\delta_{\alpha\beta}
\underbrace{\mathcal{N}^2 \frac{t+\Delta_1}{2} 
(\zeta_+^* - \zeta_-^*)
(\zeta_+^{L+1} - \zeta_-^{L+1})}_\delta
\end{align}
We know that either $\zeta_+,\zeta_-\in\mathbb{R}$ or $\zeta_+ = \zeta_-^*$, thus $\delta\in\mathbb{R}$. Then in the basis $\ket{\psi_+}, \ket{\psi_-}, \ket{\psi_+'}, \ket{\psi_-'}$, the hybridzation Hamiltonian reads
\begin{equation}
H_\text{hybr} 
=
\begin{pmatrix}
&&  -\ii \delta \\
&& & \ii \delta \\
\ii \delta \\
& -\ii\delta
\end{pmatrix} = \nu_y \sigma_z \delta
\end{equation}

We can also include time-reversal breaking terms $\Delta_2,\Delta_3\neq 0$ by means of the perturbation theory. In the surface-states subspace the matrix elements read
\begin{align}
\bra{\psi_\alpha}\Delta_2 \tau_y \sigma_0\ket{\psi_\beta}
&=
\Delta_2 
\bra{\tau_y=\alpha}\bra{\sigma_z=\alpha}
\tau_y \sigma_0
\ket{\tau_y=\beta}\ket{\sigma_z=\beta}
= \alpha\delta_{\alpha\beta}\Delta_2  \\
\bra{\psi_\alpha'}\Delta_2 \tau_y \sigma_0\ket{\psi_\beta'}
&=
\Delta_2 
\bra{\tau_y=-\alpha}\bra{\sigma_z=\alpha}
\tau_y \sigma_0
\ket{\tau_y=-\beta}\ket{\sigma_z=\beta}
= -\alpha\delta_{\alpha\beta}\Delta_2 \,.
\end{align}
Thus in the basis of surface states (ignoring overlap terms)
\begin{equation}
\Delta_2 \tau_y \sigma_0 \mapsto \Delta_2 \nu_z \sigma_z\,.
\end{equation}
Similarly, we can include add the term $\Delta_3 \tau_0\sigma_z (\cos k_x - \cos k_y)$, which becomes
\begin{equation}
\Delta_3 \tau_0\sigma_z \mapsto \Delta_3 \nu_0\sigma_z\,.
\end{equation}

Then for $q=(0,0),(\pi,\pi)$ the low-energy Hamiltonian becomes
\begin{equation}
    h = 
    \Delta_1 \nu_0 (\sigma_x k_x + \sigma_y k_y)
    + \nu_y\sigma_z \delta
    + s_q (\delta q_x^2 - \delta q_y^2)(\Delta_2 \nu_z \sigma_z
    + \Delta_3 \nu_0 \sigma_z)\,,
\end{equation}
which is the result presented in the main text. For the other two TRIMs $q=(0,\pi),(\pi,0)$, we get
\begin{equation}
    h = 
    \Delta_1 \nu_0 (\sigma_x k_x + \sigma_y k_y)
    + \nu_y\sigma_z \delta
    + 2 s_q'    
    (\Delta_2 \nu_z \sigma_z
    + \Delta_3 \nu_0 \sigma_z)
\end{equation}
with $s'_{0,\pi}=1=-s'_{\pi,0}$.

\section{Topological invariant}

We will limit our considerations to this particular form of the Hamiltonian:
\begin{equation}
H = \nu_0(\sigma_x \xi_x + \sigma_y \xi_y) + (\nu_0 \delta_1 + \nu_y\delta_2)\sigma_z\,,
\end{equation}
which captures all four atomic limits. Its eigenvalues are
\begin{equation}
E = -\epsilon_\pm,+\epsilon_{\pm}\,,\qquad
\epsilon_\pm =\sqrt{\xi^2  + (\delta_1\pm \delta_2)^2}
\end{equation}
and, denoting
\begin{equation}
\delta_1\pm\delta_2 = \epsilon_\pm\cos\theta_\pm\,,\qquad
\xi_x + \ii \xi_y = |\xi|\ee^{\ii \phi}
\,,\qquad \theta_\pm \in[0,\pi]\,.
\end{equation}
the two low-energy eigenstates read
\begin{align}
\ket{\psi_\pm} = 
\frac{1}{\sqrt{2}}
\binom{1}{\pm \ii}_\nu
\binom{-\sin(\frac{1}{2}\theta_\pm)\ee^{-\ii \phi /2}}
      {\cos(\frac{1}{2}\theta_\pm)\ee^{\ii \phi/2}}_\sigma\,.
\end{align}
We find the Berry connection
\begin{equation}
A_\pm
=
\ii\bra{\psi_\pm}\nabla\ket{\psi_\pm}
= -\frac{1}{2}\cos\theta_\pm \nabla\phi
= -\frac{1}{2}\frac{\delta_1\pm\delta_2}{\epsilon_\pm} \nabla\phi\,.
\end{equation}

At $k\in\{(0,0),(\pi,\pi)\}=\{k_1,k_2\}$, the symmetry implies $\xi_x=\xi_y=\delta_1=0$. Then $\epsilon_+ = \epsilon_- = |\delta_2|$, and $\cos\theta_\pm = \pm\mathrm{sign}(\delta_2)$. In the basis of $\ket{\psi_\pm}$, the $C_4\mathcal{T}$ opertor reads
\begin{equation}
\tilde w(k_i) = 
\begin{pmatrix}
\bra{\psi_+}C_4\mathcal{T}\ket{\psi_+} & \bra{\psi_+}C_4\mathcal{T}\ket{\psi_-} \\
\bra{\psi_-}C_4\mathcal{T}\ket{\psi_+} & \bra{\psi_-}C_4\mathcal{T}\ket{\psi_-}
\end{pmatrix}
=
\frac{1}{\sqrt{2}}
\begin{pmatrix}
0 & 1-\ii\mathrm{sign}(\delta_2) \\
1+\ii\mathrm{sign}(\delta_2) & 0
\end{pmatrix}\,,
\end{equation}
and the Pfaffian of its anti-symmetrization:
\begin{equation}
\operatorname{Pf} w(k_i) = \operatorname{Pf} \frac{\tilde w(k_i) - \tilde w^T(k_i)}{\sqrt{2}}
= -\ii \operatorname{sign}(\delta_2)
\end{equation}
The last two ingredients needed to calculate the invariant are the Wilson line and flux of the Berry curvature through the irreducible Brilloin zone (IBZ). We'll first compute the latter. The Berry curvature reads
\begin{equation}
\mathcal{F}_\pm = \nabla \times A_\pm
= 
-\frac{1}{2}\frac{\delta_1\pm\delta_2}{\epsilon_\pm} \nabla\times\nabla\phi
-\frac{1}{2}\left(\nabla\frac{\delta_1\pm\delta_2}{\epsilon_\pm}\right) \nabla\phi\,.
\end{equation}
Numerically, the first term vanishes (as it equals to a sum of delta functions). However, the contour integral picks up the singularities. Thus, when computed numerically,
\begin{equation}
\int_\text{IBZ} d^2 k \mathcal{F}_\pm -
\int_{\partial \text{IBZ}} d k \cdot A_\pm 
=
\int_\text{IBZ} d^2 k \frac{1}{2}\frac{\delta_1\pm\delta_2}{\epsilon_\pm} \nabla\times\nabla\phi\,.
\end{equation}
$\nabla\times\nabla\phi$ can only by non-zero at points where $\xi=0$. Then $\epsilon_\pm = |\delta_1 \pm \delta_2|$. If we denote by $k_j$ the points where $\xi$ vanishes (assuming $\xi$ is not identically 0, in which case the Berry flux vanishes), and by $\mu_j$ its windings around each point, then
\begin{equation}
\frac{1}{2\pi}\left(\int_\text{IBZ} d^2 k \mathcal{F}_\pm -
\int_{\partial \text{IBZ}} d k \cdot A_\pm \right)
=
\sum_{k_j\in \text{IBZ}} \frac{\mu_j}{2}\operatorname{sign}(\delta_1\pm\delta_2)|_{k=k_j}\,.
\end{equation}

Applying this result to the low-energy surface Hamiltonian $h$ with $\Delta_2=0$, we can use the fact that $\xi$ only vanishes at TRIMs. This results in
\begin{equation}
    \sum_{\alpha=\pm}\frac{1}{2\pi}\left(\int_\text{IBZ} d^2 k \mathcal{F}_\alpha -
\int_{\partial \text{IBZ}} d k \cdot A_\alpha \right)
=
-\frac{1}{2}
[ \operatorname{sign}(2 \Delta_3 + \delta_{0,\pi}) + \operatorname{sign}(2 \Delta_3 - \delta_{0,\pi})]\,.
\end{equation}
Altogether this yields the invariant
\begin{equation}
\nu
= 
1 - \operatorname{sign}(\delta_{0,0}\delta_{\pi,\pi})
- \operatorname{sign}(2 \Delta_3 + \delta_{0,\pi}) + \operatorname{sign}(2 \Delta_3 - \delta_{0,\pi})
\mod 4\,.
\end{equation}

\end{document}